# Fifth Harmonic and Five-Photon Excitation Fluorescence Multiphoton Microscopy

**JOSHUA H. MAGNUS,* LAM T. NGUYEN, ELANA G. ALEVY, TRAVIS W. SAWYER, SAMUEL D. CROSSLEY, AND KHANH KIEU**

*University of Arizona, Wyant College of Optical Sciences, 1630 E. University Blvd. Tucson, AZ, 85719*
**magnus@arizona.edu*

**We report the application of fifth harmonic and 5-photon excitation fluorescence multiphoton microscopy. This novel label-free imaging modality has been characterized with spectral and power dependence measurements. The advantages of fifth harmonic and 5-photon excitation fluorescence imaging are higher resolution and use of longer wavelength excitation sources, which may allow deeper penetration depth, while still exciting near-UV to blue fluorophores. This imaging modality may find interesting applications in biological research, materials characterization, geologic studies, and semiconductor manufacturing.**

**Introduction.** Laser scanning multiphoton microscopy (MPM) has found applications in a growing number of scientific fields for its label-free, non-destructive, high resolution 3D imaging capabilities. Two-photon excitation laser scanning microscopy has been actively used since 1990 [1] and has since become ubiquitous in biosciences for the ability to probe fluorescence at a narrow plane, in thick tissue samples, non-destructively with high resolution [2,3]. Bioscience applications such as in-vivo neuroimaging of mice [4,5], and single-cell behavior of cancerous cells [6] have been investigated with two-photon excitation fluorescence (2PEF) imaging. Similarly, three-photon excitation fluorescence (3PEF) imaging has been applied to biological applications such as ovarian cancer tissue [7] and in-vivo detection of fluorescent labeled lipids [8]. Second harmonic generation (SHG) and third harmonic generation (THG) were also found to have applications in biological samples [9–13]. Harmonic imaging (SHG, THG) has also been applied to materials characterization, such as polymer films [14] black phosphorous layered materials [15], and grain boundaries of $MoS_2$ [16]. The use of multimodal MPM incorporating all signals has been shown as a valuable tool in applications such as imaging gastric cancer for early detection [17] and in geologic studies for 3D imaging of gems and minerals [18,19]. MPM has increased in popularity in these scientific fields as it provides label-free, 3D, sub-micron resolution, nonlinear optical imaging with multiple distinct signal channels. Utilizing each of these signals (SHG, THG, 2PEF, 3PEF) can highlight different tissues, materials, structures, stresses, or interfaces, that all together gives the composite MPM image more useful information. Additional signal channels at higher order processes (>3-photon) could uncover more information in the samples imaged.

Multiphoton imaging at >3$^{rd}$ order has been demonstrated before, with four-photon excitation fluorescence imaging in a mouse brain at a pump wavelength of 1700 nm [20] However, 5-photon multiphoton imaging has not been reported so far to the best of our knowledge. Studies of fifth harmonic generation (5HG) have been theoretically reported in isotropic media, cubic centrosymmetric crystals [21,22], and experimentally demonstrated in $CaF_2$ with 2 µm pump wavelength [23] 5HG has also been utilized before in conjunction with third-harmonic signals to characterize inter-layer defects in Si wafers [24], but not using the 5HG signal as an imaging modality.

Here, we present the first use of both fifth harmonic generation (5HG) and 5-photon excitation fluorescence (5PEF) signals. Images were collected using a lab-built femtosecond Er-doped fiber laser excitation source working near 1560 nm [25] with a lab-built multiphoton microscope. We show that one can achieve higher lateral and axial resolution than conventional two and three photon imaging modalities with 5-photon imaging. 5PEF imaging has potential applications in biological research as it is possible to use laser sources at longer wavelengths and still excite fluorophores in the near-UV to blue wavelength range. Specifically, a source laser working at 1700nm [26,27] has the advantage of reducing absorption from water and less scattering allowing for deeper penetration into samples [28,29]. With 5PEF imaging, these 1700 nm sources could still excite fluorophores down to 340 nm, including important dyes such as Alexa Fluor 350 and 405, NADH, or DAPI. Though in this letter, the source will be centered at 1560 nm with detection down to 310 nm. The technique demonstrated could be applied to various systems working at other wavelengths.

Fig. 1A shows the Jablonski energy diagrams for 2$^{nd}$ and 3$^{rd}$ order multiphoton signals, (SHG, 2PEF; THG, 3PEF). Fig. 1A also shows the energy diagrams for the five-photon interactions: 5HG and 5PEF. The harmonic processes involve n-photons (at frequency $\omega$) inducing a virtual excited state and release a photon at the energy $n\hbar\omega$. Multiphoton excitation fluorescence occurs when multiple photons are absorbed at once to excite electrons to an excited state at $n\omega$, then the fluorophore relaxes and releases a photon of lower energy ($< n\hbar\omega$).

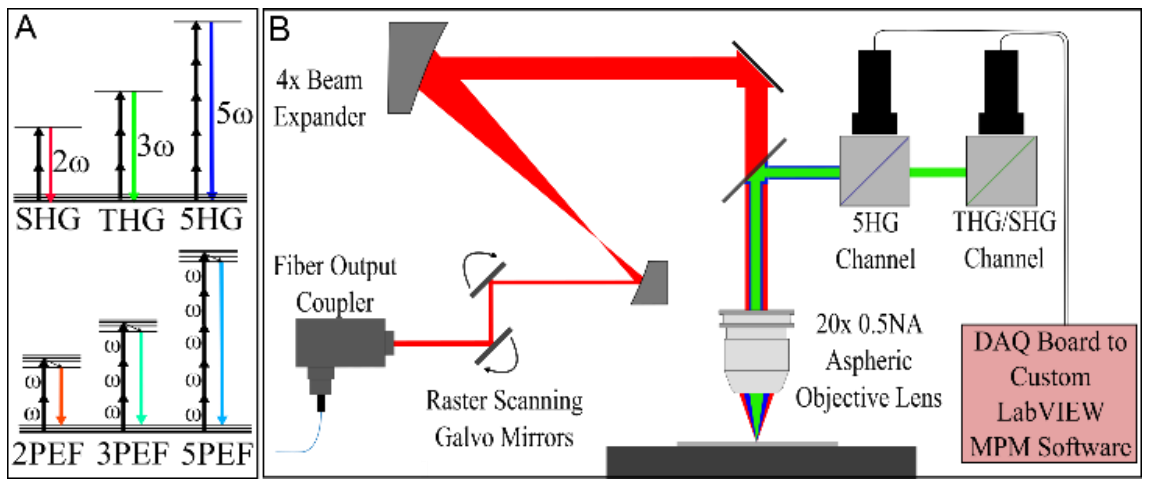

Fig. 1. (A) Jablonski diagram showing multiphoton processes second, third, and fifth harmonic generation, and two, three, and 5-photon excitation fluorescence. (B) MPM design and detection scheme for 5HG/5PEF imaging.

**Microscope and Detection.** The diagram of the MPM used for 5-photon processes imaging is shown in Fig. 1B, and the design of the MPM is described in detail in [30,31]. The laser used in this system is an Er-doped mode-locked fiber laser generating 8 nJ pulse energy and 45 fs pulse duration (40 nm bandwidth) at 8 MHz repetition rate. In this system, a 20x 0.5NA New Focus aspheric objective was used for imaging. This objective was used as it transmits >80% of the fundamental wavelength (1550 nm) onto the sample, and 18% at the fifth harmonic wavelength, 310 nm. The average power on the samples was at a maximum of 52 mW and was stepped down with ND filters as needed to avoid photodamage or photobleaching in sensitive samples. A 10x 0.5NA Nikon objective was also used for 5PEF imaging, as this objective has >10% transmission down to ~320nm and a larger FOV, but was not suitable for 5HG imaging due to lower transmission at the pump wavelength (only 24 mW reaching the sample). To isolate 5HG and 5PEF signals, a long-pass dichroic filter with cut-off at 510 nm (Semrock FF510-Di02) reflects <510 nm wavelengths toward the PMT (Hamamatsu H10721-110), filtering out >93% of the light from 515 nm to 950 nm. To further isolate the 5HG-5PEF wavelength range, a UV colored glass filter (Thorlabs FGUV11) was used with transmission averaging 80% within 300 nm - 370 nm. The UV filter also importantly extinguishes the third harmonic wavelength to OD4 (main source of cross-signal detection). A custom 310 nm bandpass filter (60% transmission, 8nm bandwidth) was also used to isolate 5HG signal. The system could be improved with a high NA objective with optimal transmission in both the 5HG/5PEF and pump wavelength ranges to increase signal strength. For detection of other standard MPM signals (SHG, THG, 2PEF, 3PEF) appropriate dichroic mirrors and bandpass filters were used [31].

**Results.** Fifth harmonic signals were observed in semiconductor materials (GaAs, Si, and GaN), and 5-photon fluorescence signals were found to be common in samples that have fluorescence emission in the 310 nm-370 nm wavelength band. Samples imaged with 5HG & 5PEF in this report include Gallium Arsenide computer processor chips, a geologic specimen, and human tissue samples. This indicates that 5HG/5PEF are quite prevalent in many materials. Furthermore, the pulse energy used for excitation is only < 10 nJ which is below the damage threshold for most samples.

As shown in Fig. 2A-B, the power dependence of the 5HG and 5PEF signals from GaAs and Intel Pentium CPU die (Fig. 4) were found to have a slope of 4.9 ± 1.3 and 5.5 ± 1.7 in a log-log plot, respectively. The slopes agree with the expected fifth-order power dependence, though the fitted slopes have uncertainties which leave the possibility for four or even six photon processes. Spectral measurements were also performed to establish without doubt the nature of the five-photon interaction. The results of these measurements are shown in Fig. 2C-D. Here, an Ocean Optics spectrometer (USB4000) was used to collect the 5HG spectrum from the GaAs sample and the 5PEF signal from the Intel processor. The 5HG signal is 10 dB above the noise floor, centered at 310 nm, with a bandwidth of 4 nm. The 5HG spectrum was collected with a 560 nm dichroic and a Thorlabs FG-UV05 color glass filter (transmission of 97% at 310 nm, and 0.02% at 520 nm), to highlight both the 5HG and THG signals. The 5PEF signal was found to be broadband in the range of 327 nm – 420 nm.

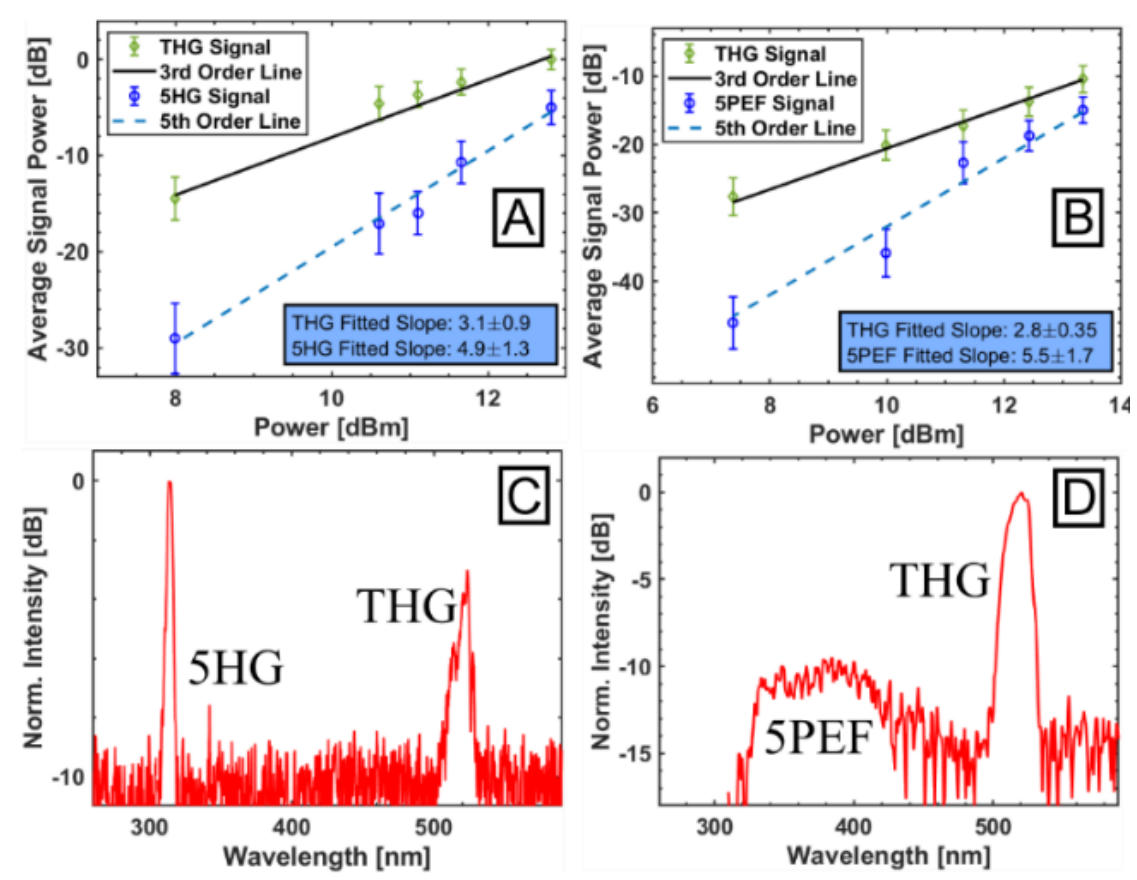

Fig. 2. Power dependence and spectral measurements of THG and 5HG (A, C) and 5PEF (B, D) signals in GaAs and a Pentium 4 CPU chip.

Next, we show the improved resolution of 5-photon processes in relation to two and three photon imaging modalities. The theoretical diffraction limited lateral and axial resolution for multiphoton imaging has been previously established [2]:

$$r_{xy} = \begin{cases} \dfrac{0.533\,\lambda}{\sqrt{m}\,NA}, & NA < 0.7 \\ \dfrac{0.541\,\lambda}{\sqrt{m}\,NA^{0.91}}, & NA \geq 0.7 \end{cases} \qquad r_z = \frac{0.886\lambda}{\sqrt{m}}\left[\frac{1}{n - \sqrt{n^2 - NA^2}}\right]$$

The lateral resolution (full-width half max value $r_{xy}$) of an MPM is limited by the excitation wavelength ($\lambda$), numerical aperture of the objective ($NA$), and the order of the multiphoton process ($m$). As the order increases, the theoretical resolution increases. In this case, using a 20x 0.5NA C-coated New Focus aspheric objective, the theoretical lateral (and axial) resolution for second, third, and fifth harmonic processes are 1.17 µm (7.25 µm), 0.952 µm (5.92 µm), and 0.738 µm (4.58 µm), respectively. For samples where 5-photon signals are readily apparent, the lateral and axial resolution can be improved by 22%, in theory, compared to the three-photon process.

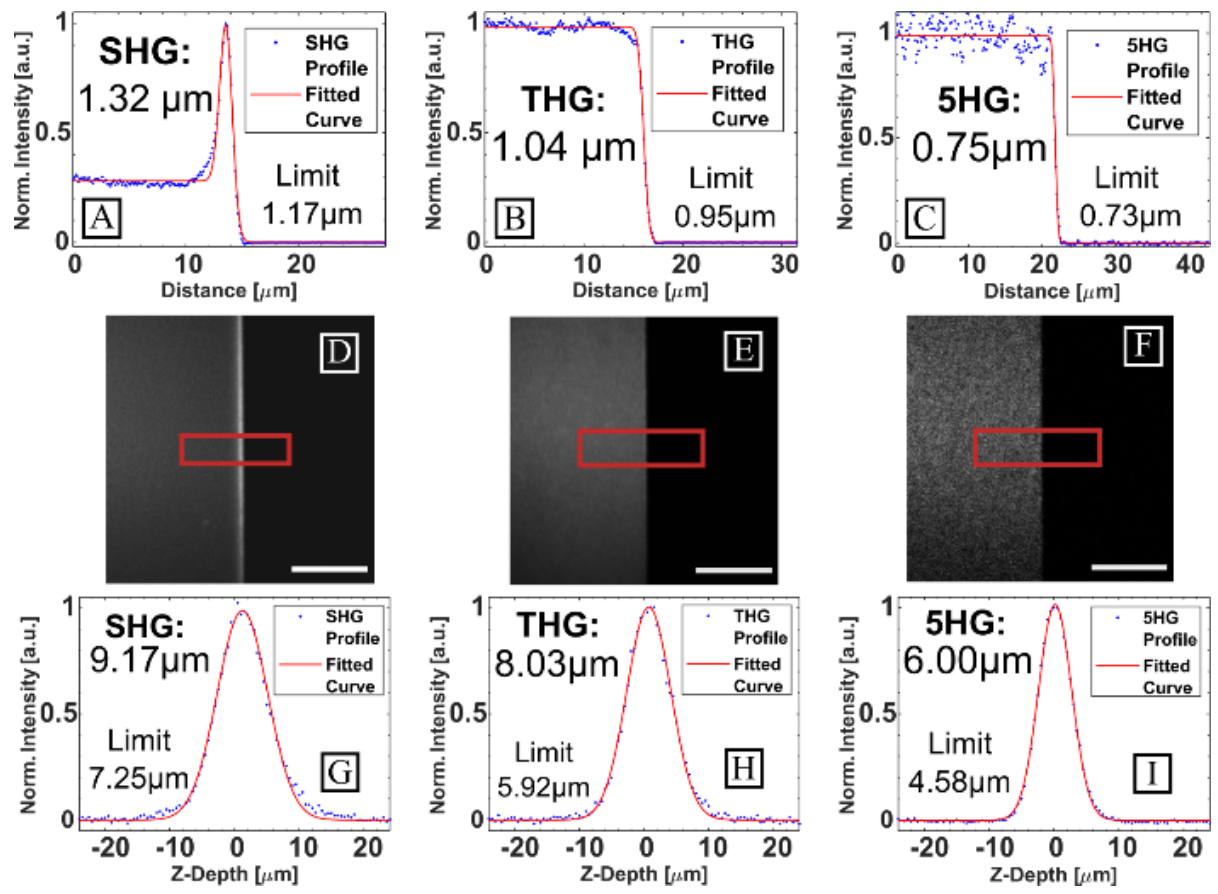

Fig. 3. Nonlinear knife-edge lateral resolution measurements of GaAs for SHG, THG, and 5HG signals shown in (A-C). (A) has a peak due to broken symmetry at the edge enabling SHG. Images of GaAs edge shown in (D-F) with 20 µm scale bars. Axial resolution measured from z-stack of GaAs surface profile and fitted Gaussian function are shown in panels (G-I).

The lateral resolution was measured using the nonlinear knife-edge technique [32]. SHG, THG and 5HG images were obtained at the edge of a GaAs wafer (Fig. 3D-F), and the profile of the knife-edge is characterized as described in [32], and shown in Fig. 3A-C. The measured resolution for second, third, and fifth harmonic processes were calculated as 1.32 µm, 1.04 µm, and 0.75 µm. To measure the axial resolution, a z-stack of images was taken at the surface of the GaAs. The axial profile data was collected from the intensity data of the GaAs z-stack, and assuming a Gaussian profile, fitting was performed to characterize the axial resolution. To accurately measure the axial resolution, the power was stopped down with ND filters (0.4 ND for 5HG, 1.4 ND for SHG/THG) to avoid saturation effects broadening the z-profile. The SHG, THG, and 5HG axial resolution were measured to be 9.08 µm, 8.0 µm, and 6.02 µm, respectively. 5HG was measured to have a 27% better lateral resolution and a 24% better axial resolution compared to the THG results. Our resolution measurements are in good agreement with the theoretical estimates shown above.

To demonstrate a practical application of 5-photon imaging modalities, we have imaged an Intel Pentium 4 CPU by removing the die from the package and polishing the surface to remove the epoxy and top metal layers. This exposed the internal structure of the CPU, which can be mapped out and characterized with MPM imaging (CPUs are known to have sub-micron structures). In Fig. 4A-C, the images are composites of two channels: THG (green) and 5PEF (blue). The strongest fluorescent signal was observed in the 5PEF region (Fig. 2D), while there is another separate 2PEF signal (Semrock FF01-850/10 bandpass filter was used) which apparently also maps out the same structure as the 5PEF signal in the CPU. The 2PEF and 5PEF images are compared in Fig. 4D-E, and a line profile is plotted from each image. The full width-half max of a line was measured for the 2PEF and 5PEF channels. The 2PEF channel measured a width of 2.6 µm, while the 5PEF channel measured a FWHM of 2.2 µm. The difference in widths is due to the convolution of the object (~2.2 µm size) and the 2PEF/5PEF point spread functions, demonstrating the increased resolution of the 5-photon imaging process. The increased resolution of the 5PEF channel also shows better contrast of the smaller structures in the image seen in Fig. 4D-E. This produces a clear advantage for the 5-photon imaging modality even when other fluorescent channels may image the same structures. This imaging technique could be applied to semiconductor technology at the manufacturing level to rapidly identify any defects or characterize layers as they are deposited to validate manufacturing processes.

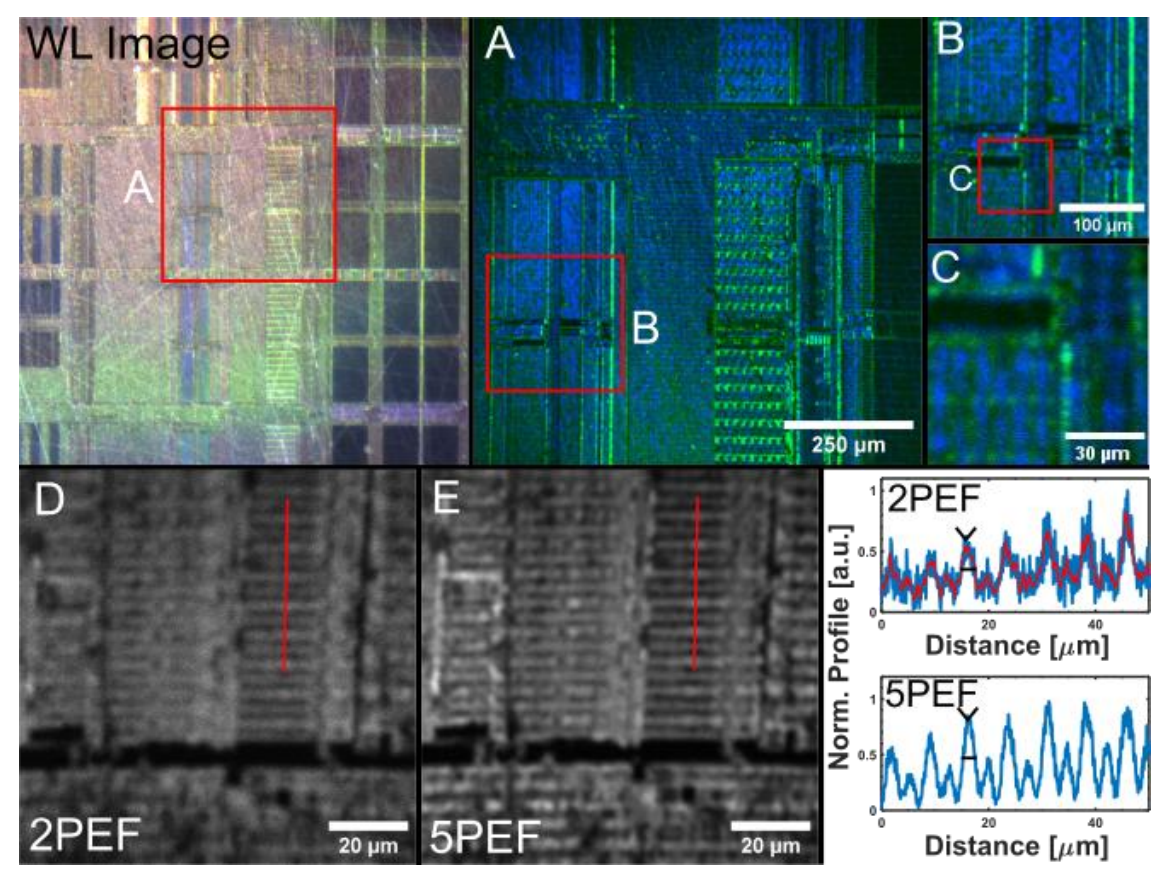

Fig. 4. Intel Pentium 4 CPU. White light (WL) images taken with an Olympus SZ60 microscope, A-C: Images of CPU THG-green; 5PEF-blue. D-E: 2PEF/5PEF images of CPU. Line profiles (red lines in D&E) show feature widths of 2.2 µm and 2.6 µm for 5PEF/2PEF (indicated feature with V). Images acquired with the 10x 0.5 NA objective, 24 mW on the sample and pixel dwell time of 10 µs.

We have recently shown that MPM is useful in imaging gems and minerals [18,19] It turns out that 5-photon processes can be present in gems and minerals as well. Fig. 5A-C shows a 3D rendering of a zoned fluorite sample, with the signals found to be SHG (red), THG (green), and 5PEF (blue). A set of images were taken at depth spanning 810 microns, spaced 3 microns apart. 5PEF signal was found in features/inclusions in the fluorite hundreds of microns below the surface. The structure mapped out with the 5PEF channel was unique to this signal channel and was not present from other MPM channels. Therefore, when acquiring MPM images using a pump wavelength at 1550 nm, the only way to map out these structures is to observe at 5-photon emission wavelengths.

5-photon signal was also found in human pancreas samples. The sample (IRB #0600000609) imaged was an unstained 7-micron thick fixed pancreatic cancer tissue sample. The MPM image in Fig. 6 has four channels SHG (red), THG (green), 3PEF (cyan), and 5PEF (blue). The 4.2 mm by 4.2 mm image of the pancreas sample was created by stitching a grid of 20x20 images. Select locations from the mosaic image are shown in Fig. 6A-D. Significant 5PEF signal is seen in a feature (possibly a transverse blood vessel) highlighted in Fig. 6A, as well as SHG around the structure indicating collagen. In Fig. 6B-C, 5PEF signal is also observed in the fine details of the structure (possibly islets), seen in more detail in Fig. 6D. This fine structure is only revealed in the 5PEF channel, once again, collecting signal at the 5-photon wavelengths is the only way to observe these structures at this pump wavelength.

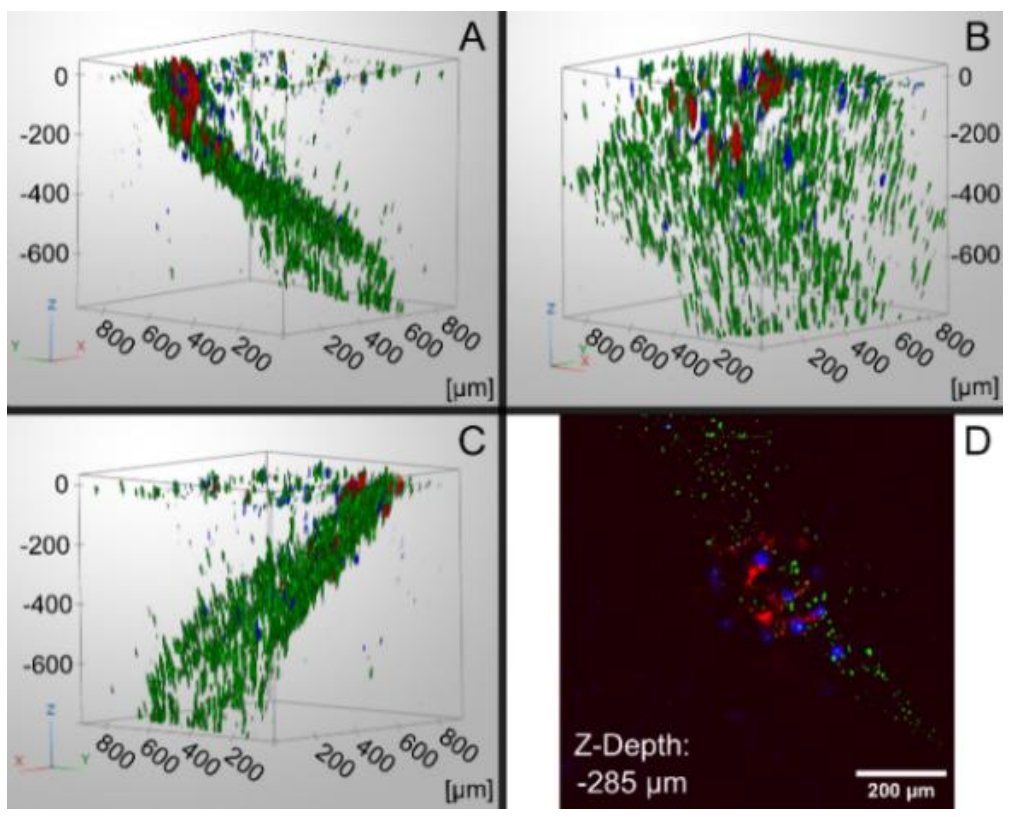

Fig. 5. Green-THG; Red-SHG; Blue-5PEF. A-C: 3D rendering of structure in zoned fluorite sample, shown in successive angles (axis units in microns). D: Single frame from the image stack, at a depth of -285 µm below the surface of the sample. Images acquired with a 10x 0.5 NA objective, 24 mW on the sample and 15 µs dwell time.

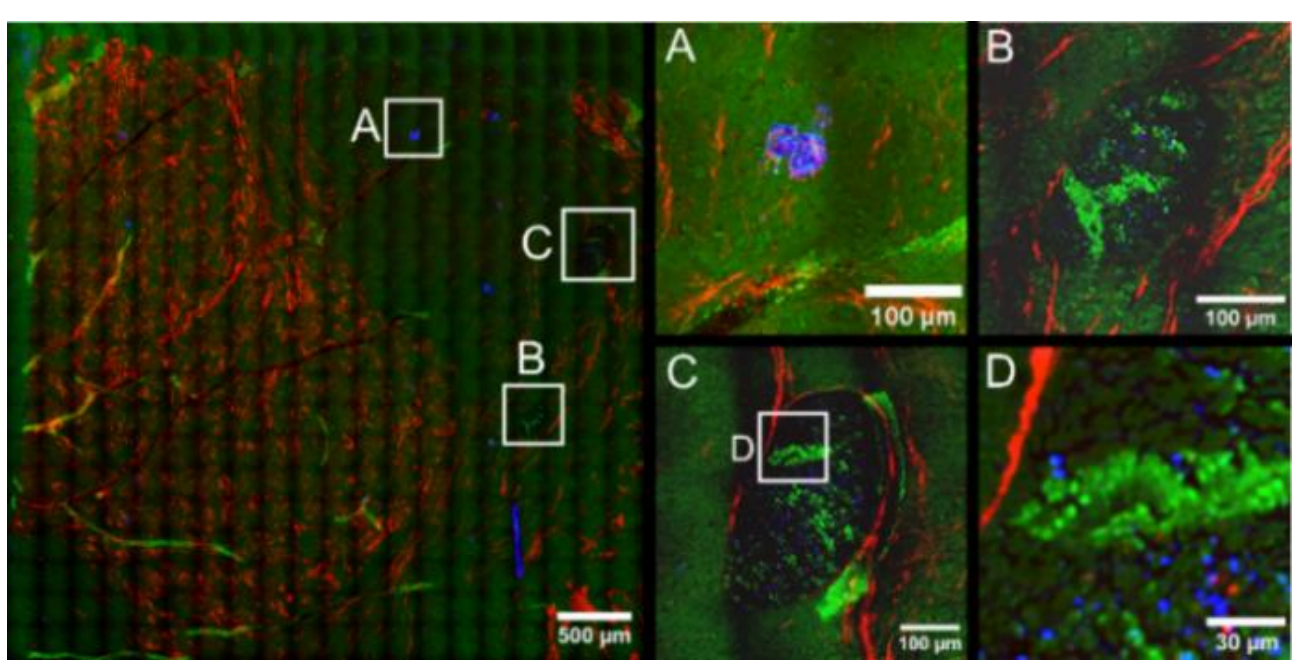

Fig. 6. Human pancreas tissue sample. SHG – Red; THG – Green; 3PEF – Cyan, 5PEF – Blue. Area A, B, C are highlighted in the main mosaic image, and area D is a cropped section of area C. Images acquired with 20x 0.5 NA aspheric objective with 52 mW on the sample and a pixel dwell time of 10 µs.

**Conclusion.** 5HG and 5PEF multiphoton microscopy has been shown to have measurably better lateral and axial resolution and with significant fluorescence signals observed in various samples. This novel imaging modality has been shown to be appropriate and even favorable for different applications in several scientific fields. 5-photon imaging may have application in the semiconductor industry. This imaging modality in conjunction with other multiphoton signals could be a useful tool for validation and defect inspection in electronic circuits. Geologic studies using MPM as a tool for 3D non-destructive characterization could benefit from this imaging modality, as 5PEF imaging enables the use of longer excitation wavelengths that scatter less, thus can be used for deeper inspection of samples without compromising on lower-wavelength fluorescent signals. Similarly, in biological research, longer wavelengths can be advantageous to image deeper into tissue samples, as well as increased resolution with 5-photon imaging. This imaging modality can expand the capabilities of any existing multiphoton microscope.

**Disclosures.** The authors declare no conflicts of interest.

**Data availability.** Data underlying the results presented in this paper are not publicly available at this time but may be obtained from the authors upon reasonable request.